\documentstyle[epsfig,aps]{revtex}

\def\gsim{ \,\, \vcenter{\hbox{$\buildrel{\displaystyle >}\over\sim$}}
 \,\,}
\def\lton{ \,\, \vcenter{\hbox{$\buildrel{\displaystyle <}\over\sim$}}
 \,\,}

\newcommand{\be}{\begin{eqnarray}}
\newcommand{\ee}{\end{eqnarray}}
\newcommand{\non}{\nonumber\\}
\newcommand{\bea}{\begin{eqnarray}}
\newcommand{\eea}{\end{eqnarray}}
\newcommand{\PolL}{{\cal P}_\Lambda}

\begin{document}

\thispagestyle{empty}
\title{\bf Polarized Hyperons from $p\, A$ Scattering in the Gluon Saturation
Regime}

\author
{
Dani\"el~Boer$^a$ and Adrian~Dumitru$^{b,c}$\\
 {\small\it $^a$Department of Physics and Astronomy, Vrije Universiteit, De
Boelelaan 1081, NL-1081 HV Amsterdam, The Netherlands}\\
 {\small\it $^b$Institut f\"ur Theoretische Physik, Universit\"at Frankfurt,
Postfach 111932, D-60054 Frankfurt am Main, Germany}\\
 {\small\it $^c$Physics Department, Brookhaven National Lab, Upton,
            NY 11973, USA}\\
}

\maketitle

\begin{abstract}
We study the production of transversely polarized $\Lambda$ hyperons in
high-energy collisions of protons with large nuclei. The large gluon density
of the target at saturation provides an intrinsic semi-hard scale which
should naturally allow for a weak-coupling QCD description of the process
in terms of a convolution of the quark distribution of the proton with
the elementary quark-nucleus scattering cross section (resummed to all
twists) and a fragmentation function. In this case of transversely polarized
$\Lambda$ production we employ a so-called polarizing fragmentation 
function, which is an odd function of the transverse momentum of the $\Lambda$
relative to the fragmenting quark. Due to this $k_t$-odd nature, the
resulting $\Lambda$ polarization is essentially proportional to the
{\em derivative\/} of the quark-nucleus cross section with respect to
transverse momentum, which peaks near the saturation momentum scale.
Such processes might therefore provide generic signatures for
high parton density effects and for the approach to the ``black-body''
(unitarity) limit of hadronic scattering.
\end{abstract}

\vspace*{1cm}

It has been known for over 25 years that $\Lambda$'s produced in
collisions of unpolarized hadrons exhibit polarization
perpendicular to the production plane. 
As of yet, such data are not available for very high energies where one
expects that hadronic cross sections are close to their geometrical values
(the ``black body limit''). However, the BNL-RHIC collider will soon collide
protons and deuterons on gold nuclei at energies of $\sim200$~GeV in the
nucleon-nucleon center of mass frame; later on, much higher
energies will be accessible at the CERN-LHC. In this letter, we demonstrate 
that the polarization of $\Lambda$ hyperons produced in the forward region
in high-energy collisions of protons and heavy nuclei may generically
be a sensitive probe of high-density effects and gluon saturation in the
target.

The wave function of a hadron (or nucleus) boosted to large rapidity exhibits
a large number of gluons at small $x$, which is the fraction of the light-cone
momentum carried by the gluon. The density of gluons is expected to saturate
when it becomes, parametrically, of the order of the inverse QCD
coupling constant $\alpha_s$~\cite{mq}. The parton density at
saturation is denoted by $Q_s^2$, the so-called saturation momentum. This
provides an intrinsic momentum scale~\cite{sat} which grows with atomic number
and with rapidity because more gluons can be radiated in the initial state
when phase space is big.
For sufficiently high energies and/or large nuclei, the saturation momentum
$Q_s$ can become much larger than $\Lambda_{\rm QCD}$, such that weak
coupling methods are applicable.

Forward $\Lambda$ production in $pA$ collisions
is dominated by high-$x$ quarks from the proton traversing the
high gluon density region of the heavy nucleus.
The quarks typically experience
interactions with momentum transfers of the order of the saturation momentum.
Thus, for large gluon densities in the target, such that the saturation
momentum is in the perturbative regime, $Q_s\gsim1$~GeV, the coherence of
the projectile is lost, and the scattered quarks (having an average 
transverse momentum proportional to $Q_s$) fragment independently~\cite{dgs}.
While nonperturbative constituent-quark and diquark scattering and
hadronization models~\cite{nonpert} have been employed to understand hyperon
polarization in collisions of protons with dilute targets, we expect that in
the high-energy limit the presence of the intrinsic semi-hard
scale $Q_s$ should naturally allow for a weak-coupling QCD description of the
process. One can thus calculate the cross
section for $q A$ scattering in this kinematical domain within 
pQCD~\cite{djm2}, and the deflected, outgoing quark will subsequently 
fragment into hadrons, which is described by a fragmentation function. 

\begin{figure}[htp]
\centerline{\hbox{\epsfig{figure=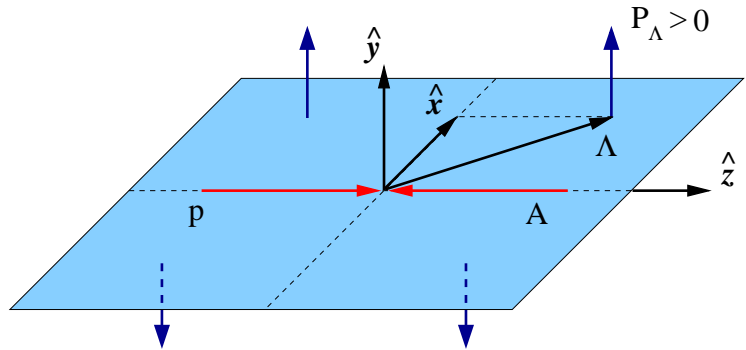,height=5cm}}}
\mbox{}\\
\caption{Kinematics of the $p \, A \to \Lambda \, X$ process. The direction 
of positive $\Lambda$ polarization is indicated for each quadrant in the 
$\Lambda$ production plane.}
\label{fig0}
\end{figure}

In order to explain the transverse $\Lambda$ polarization in unpolarized
hadron collisions within such a factorized pQCD description, it has been 
suggested that unpolarized quarks can 
fragment into transversely polarized hadrons, for instance $\Lambda$ 
hyperons. The associated probability \cite{Mulders:1995dh,ABDM1} is 
described by a 
so-called polarizing fragmentation function, sometimes also called Sivers
(effect) fragmentation function. Its main properties are that it is an odd
function of the transverse momentum relative to the
quark, $\vec{k}_t$, and that the $\Lambda$ polarization is 
orthogonal to $\vec{k}_t$, because of parity invariance.
The polarizing fragmentation function is defined as 
\cite{ABDM1}\footnote{Another commonly
used notation for the polarizing fragmentation function is $D_{1T}^\perp$, but
with a slightly different definition \cite{Mulders:1995dh}.}: 
\be
\Delta^N D_{h^\uparrow/q}(z,\vec{k}_t) \equiv
\hat D_{h^\uparrow/q}(z,\vec{k}_t) - 
\hat D_{h^\downarrow/q}(z,\vec{k}_t)
= \hat D_{h^\uparrow/q}(z,\vec{k}_t)-
\hat D_{h^\uparrow/q}(z,-\vec{k}_t), 
\ee 
and denotes the difference between the densities  
$\hat D_{h^\uparrow\!/q}(z, \vec{k}_t)$ and  
$\hat D_{h^\downarrow\!/q}(z,\vec{k}_t)$ 
of spin-1/2 hadrons $h$, with longitudinal momentum fraction $z$, transverse  
momentum $\vec{k}_t$ and transverse polarization $\uparrow$ or  
$\downarrow$, in a jet originating from the fragmentation of an  
unpolarized parton $q$. Clearly, this $k_t$-odd function vanishes when
integrated over transverse momentum and also when the transverse momentum 
and the transverse spin are parallel. In order to set the sign convention for 
the $\Lambda$ polarization we define
\be
\Delta^N D_{h^\uparrow/q}(z,\vec{k}_t) \equiv 
\Delta^N D_{h^\uparrow/q}(z,|\vec{k}_t|) \;  
\frac{\vec{P}_h \cdot (\vec{q} \times \vec{k}_t)}{|\vec{q} \times 
\vec{k}_t|},  
\ee
where $\vec{q}$ is the momentum of the unpolarized quark that fragments and 
$\vec{P}_h$ is the direction of the polarization vector of the hadron $h$ 
(the $\uparrow$
direction). Fig.~\ref{fig0} shows the kinematics of the process under
consideration and indicates the direction of positive $\Lambda$ polarization
for each quadrant in the $\Lambda$ production plane.

It should be emphasized that such a nonzero probability difference 
$\Delta^N D_{h^\uparrow/q}(z,\vec{k}_t)$ is allowed by both parity and time
reversal invariance. Generally it is expected to occur due to final state
interactions in the fragmentation process, where the direction of the
transverse momentum yields an oriented orbital angular momentum compensated 
by the transverse spin of the final observed hadron. This polarizing
fragmentation function is the analogue of the so-called Sivers effect for
parton distribution functions \cite{s90}, which yields different probabilities
of finding an unpolarized quark in a transversely polarized hadron,
depending on the directions of the transverse spin of the hadron and the
transverse momentum of the quark. The Sivers effect can lead to single spin
asymmetries, for instance in $p^\uparrow \, p \to \pi \, X$, a process for
which such (large) asymmetries have been observed in several experiments. 

Recently, such a single spin asymmetry in $e \, p^\uparrow \to e' \, \pi \, X$
has been calculated in a one-gluon exchange model~\cite{BHS}. Shortly
afterwards it was understood \cite{Collins-02} 
as providing a model for the Sivers effect distribution function. 
A similar calculation has recently been performed by Metz \cite{Metz} for
the production of polarized spin-1/2 hadrons in unpolarized scattering, which
can be viewed as providing a model for the polarizing fragmentation function. 
Here we will not employ such a model calculation, but rather use a 
parametrization for the polarizing fragmentation functions obtained from a fit
to data \cite{ABDM1}. However, these model calculations do demonstrate that 
nonzero Sivers effect functions can arise in principle.

Due to the $k_t$-odd nature of the polarizing fragmentation function it is 
accompanied by a different part of the partonic cross section (essentially 
the first derivative w.r.t.\ $k_t$) compared to the ordinary,
unpolarized $\Lambda$ fragmentation function, which is $k_t$-even. 
The characteristics of the resulting $\Lambda$ polarization will turn out to 
be rather 
different from presently available data for hadronic collisions at moderately 
high energies and with ``dilute'' targets. These data show a $\Lambda$ 
polarization that increases approximately linearly as a function of the
transverse momentum $l_t$ of the $\Lambda$, up 
to $l_t \sim 1$ GeV/c, after which it becomes flat, up to the highest 
measured $l_t$ values: $l_t \sim 4$ GeV/c. No indication of a decrease 
at these high $l_t$ values has been observed. Furthermore, the polarization 
increases with the longitudinal momentum fraction $\xi$ and is to
a large extent $\sqrt{s}$ independent. These features do not
change with increasing $A$ \cite{Abe:su,Heller:ia,Panagiotou:1989sv}. 
The only $A$ dependence observed is a slight overall suppression of 
the $\Lambda$ polarization for large A and higher energies. 
For Cu and Pb fixed targets, probed with a 400 GeV/c proton beam 
\cite{Heller:ia,Panagiotou:1989sv}, the magnitude of the
polarization is about 30~\% lower than for light nuclei. This effect
is usually attributed to secondary $\Lambda$ production through $\pi^- N$
interactions~\cite{Panagiotou:1989sv}. 
The slight suppression shows no evidence for a dependence on 
$l_t$ in the investigated range 
0.9 $ < l_t < $ 2.6 GeV/c, albeit with rather low statistical accuracy. 
It is clear that this data on heavy nuclei is not in the kinematic region
where saturation is expected to play a dominant role and 
the main differences to the results presented below are that in the
saturation regime the transverse $\Lambda$ polarization {\em will\/} 
depend on the collision energy and no plateau region is expected. 

We shall now present our calculation of $\Lambda$ polarization in the
gluon saturation regime, following ref.\ \cite{ABDM1} regarding the
treatment of the polarizing fragmentation functions. 

As mentioned above, in the calculation of the $q A$ cross section one is
dealing with small coupling if the target nucleus is very dense; however, the
well known leading-twist pQCD can not be used when the density of 
gluons is large. Rather,
scattering amplitudes have to be resummed to all orders in $\alpha_s^2$
times the density. When the target is
probed at a scale $\lton Q_s$, scattering cross sections approach the
geometrical ``black body'' limit, while for momentum transfer far above
$Q_s$ the target appears dilute and cross sections are 
approximately determined by the known leading-twist pQCD expressions.

At high energies, and in the eikonal
approximation, the transverse
momentum distribution of quarks is essentially given by the 
correlation function of two Wilson lines $V$ running along the light-cone 
at transverse separation $r_t$
(in the amplitude and its complex conjugate),
\be
\sigma^{qA} = \int \frac{d^2q_t dq^+}{(2\pi)^2} \delta(q^+ - p^+)
\left<\frac{1}{N_c}\,{\rm tr}~\left| 
\int d^2 z_t \, e^{i\vec{q}_t\cdot \vec{z}_t} \left[
V(z_t)-1\right] \right|^2\right>~.
\ee
Here, $P^+$ is the large light-cone component of the momentum of
the incident proton, and that of the
incoming quark is $p^+=x P^+$ ($q^+$ for the outgoing quark).
The correlator of Wilson lines
has to be evaluated in the background field of the target nucleus.
A relatively simple closed expression can be obtained~\cite{djm2} in the
``Color Glass Condensate'' model of the small-$x$ gluon distribution of the
dense target~\cite{sat}.
In that model, the small-$x$ gluons are described as a
classical non-abelian Yang-Mills field arising from a stochastic source of
color charge on the light-cone which is averaged over with a Gaussian
distribution. The quark $q_t$ distribution is then given by~\cite{djm2}
\be
q^+ \frac{d\sigma^{qA}}{dq^+d^2q_t d^2b} &=& \frac{q^+}{P^+} \, \delta\left(
\frac{p^+ - q^+}{P^+}\right) \frac{1}{(2\pi)^2}\, C(q_t)~,\non
C(q_t) &=& \int d^2r_t \, e^{i\vec{q}_t\cdot \vec{r}_t}
\left\{\exp\left[-2Q_s^2 \int \frac{d^2 p_t}{(2\pi)^2}\frac{1}{p_t^4}
\left(1-\exp(i \vec{p}_t \cdot \vec{r}_t)\right)\right]
-2\exp\left[-Q_s^2
\int \frac{d^2 p_t}{(2\pi)^2}\frac{1}{p_t^4}\right]
+1\right\}~.  \label{qAXsec}
\ee
This expression is valid to leading order in $\alpha_s$ (tree level), but to 
all orders in $Q_s$ since it resums any number of scatterings of the impinging
quark in the strong field of the nucleus. The saturation momentum $Q_s$, as
introduced in eq.~(\ref{qAXsec}), is related to $\chi$, the total
color charge density squared (per unit area) from the nucleus integrated
up to the rapidity $y$ of the probe (i.e.\ the projectile quark), by
\be
Q_s^2 = 4\pi^2\alpha_s^2 \frac{N_c^2-1}{N_c} \chi~.
\ee
In the low-density limit, $\chi$ is related to the ordinary leading-twist
gluon distribution function of the nucleus, see for example~\cite{gy_mcl}.
From BFKL evolution, $Q_s^2$ evolves as $\sim \exp(\lambda y)\sim x^\lambda$,
with the intercept $\lambda\simeq0.3$~\cite{BFKL}. Thus, if $Q_s^2\simeq
10$~GeV$^2$ at the proton beam rapidity (i.e.\ $x=1$) and for $A\simeq200$
targets~\cite{djm2}, then $Q_s\simeq3$~GeV at $x=0.6$, decreasing to
$Q_s\simeq2$~GeV at $x=0.05$; furthermore, assuming $Q_s^2\sim A^{1/3}$
scaling, then at $x=0.6$, $Q_s$ drops from 3~GeV to 2~GeV when the atomic
number $A$ of the target decreases from $200\to 20$. It is clear therefore
that in order to be sensitive to high-density effects, experimentally one
should study high-energy $p\, A$ collisions in the forward region (where the
polarization is largest anyway, see below) and with large target nuclei, and
then compare to $p\, p$ collisions.
Below, we shall focus on polarized $\Lambda$ production in a relatively small
rapidity interval in the forward region, and so take $Q_s$ as a constant of
order $2-3$~GeV.

The integrals over $p_t$ in eq.~(\ref{qAXsec}) are
cut off in the infrared by some cutoff $\Lambda$,
which we assume is of order $\Lambda_{\rm QCD}$.
We denote the momentum of the produced $\Lambda$ by $\vec{l}=z\vec{q}+\vec{k}$,
with $\vec{k}$ the transverse momentum relative to the fragmenting quark.
Assuming parity conservation in the hadronization process, only the component
of $\vec{k}$ in the production plane contributes to the polarization
$\PolL$, therefore in order to simplify the kinematics we choose $k_y=0$ 
as was done in 
Ref.\ \cite{ABDM1}. For forward kinematics, $q^+\gg q_t$, one then finds
$zq_t\simeq l_t-k_t$. The polarized cross section is given by
\bea
\PolL(l_t,\xi)\, \xi \frac{d\sigma}{d\xi d^2l_t d^2b} &=&
  \int d\left(\frac{q^+}{P^+}\right) \int\frac{dz}{z^2} f_{q/p}(x,
Q^2) \int \frac{d^2k_t}{(2\pi)^2} \Delta^N 
D_{\Lambda^\uparrow/q}(z,Q^2,\vec{k}_t)
\frac{q^+}{P^+}\, \delta\left(\frac{q^+}{P^+}-x\right)\, C(q_t) \\
&=& \int\frac{dz}{z^2} f_{q/p}(x,
   Q^2) \int \frac{d^2k_t}{(2\pi)^2} \Delta^N 
D_{\Lambda^\uparrow/q}(z,Q^2,\vec{k}_t)\, x \, C(q_t)\\
&=& \int\limits_\xi^1 dx
\frac{x}{\xi} f_{q/p}(x,Q^2) \int \frac{d^2k_t}{(2\pi)^2}
\Delta^N D_{\Lambda^\uparrow/q}\left(\frac{\xi}{x},Q^2,\vec{k}_t\right)\, 
C(q_t)~,
\eea
where $\xi=l_z/P_z\simeq xz$ is the longitudinal
momentum fraction carried by the $\Lambda$.
We assume that $\Delta^N D_{\Lambda^\uparrow/q}(z,
\vec{k}_t)$ is strongly peaked around an average $\vec{k}^0_t$ lying in the
production plane, such that~\cite{ABDM1}
\be
\int d^2k_t\,  \Delta^N D_{\Lambda^\uparrow/q}(z,Q^2,\vec{k}_t) F(\vec{k}_t)
\simeq \Delta^N D_{\Lambda^\uparrow/q}(z,Q^2) \left[ F(k^0_t)-F(-k^0_t)\right]~.
\ee
Note that $k^0_t$ is a function of $z$, see below. Alternatively, one could
consider Gaussian distributions over $k_t$~\cite{ABDM2}, though the above
simplified treatment is sufficient for our purposes.

Considering the unpolarized cross section, we can safely neglect
$k_t$-smearing from hadronization, which is of order $\Lambda_{\rm QCD}$,
while the quarks are typically scattered to much larger transverse momenta,
namely of order $Q_s$. 
In our numerical results shown below we also include the
contribution from anti-quarks and gluons to the unpolarized cross section,
although this is a small correction for $\xi \gsim 0.1$.
For the polarizing fragmentation functions, only contributions from $u$, $d$,
and $s$ quarks (the valence quarks of the $\Lambda$) are
considered~\cite{ABDM1}. Thus, we obtain
\be \label{Pol}
\PolL(l_t,\xi) = \frac
{\int_\xi^1 dx \, {x} f_{q/p}(x,Q^2) 
 \Delta^N D_{\Lambda^\uparrow/q}\left(\frac{\xi}{x},Q^2\right) 
 \left[ C\left(\frac{x}{\xi}(l_t-k^0_t)\right) - 
        C\left(\frac{x}{\xi}(l_t+k^0_t)\right) \right]
}
{\int_\xi^1 dx \, {x} f_{q/p}(x,Q^2) 
 D_{\Lambda/q}\left(\frac{\xi}{x},Q^2\right) 
 C\left(\frac{x}{\xi}l_t\right)
}~.
\ee
The factorization scale is chosen to
be the saturation momentum of the dense nucleus, $Q^2=Q_s^2$. A
parametrization for $\Delta^N D_{\Lambda^\uparrow/q}(z)$ in terms of the
unpolarized fragmentation function $D_{\Lambda/q}(z)$ was given 
in ref.~\cite{ABDM1}. It was obtained by performing a fit to
available $p A \to \Lambda^\uparrow X$ data (for light nuclei only), where 
the transverse momentum $l_t$ was required to be larger than
1 GeV/c, in order to justify the application of a factorized expression and 
of pQCD for the partonic cross section. Although doubts have arisen about 
the applicability of pQCD in the kinematic region covered by
the available data \cite{ABDM2}, the 
resulting functions do exhibit reasonable features. Here, we shall employ
those functions as an Ansatz to
investigate the dependence of the $\Lambda$ polarization 
on the saturation momentum $Q_s$, which turns out not to depend on the
detailed parameterization of the polarizing fragmentation functions. Rather,
it is the $k_t$-odd structure (and the fact that it is peaked around an 
average nonzero transverse momentum) that is responsible for the dependence. 
Of course, future parameterizations can be easily implemented. 

To be explicit, we use
\be
\Delta^N D_{\Lambda^\uparrow/q}(z,Q^2) \equiv 
N_q \, z^{c_q}\, (1-z)^{d_q} \, \frac {D_{\Lambda/q}(z,Q^2)}2~,
\label{polFFparam}
\ee
where
\be
N_u = N_d = - 28.13, \quad N_s = 57.53, \quad 
c_q = 11.64, \quad d_q = 1.23.
\ee
The average transverse momentum $k_t^0$ acquired in the fragmentation is
parameterized as 
\be
k_t^0 = 0.66 \, z^{0.37} \, (1-z)^{0.50} \, \mbox{{\rm GeV}}/c.
\ee
For the unpolarized fragmentation function $D_{\Lambda/q}$ in eq.\
(\ref{polFFparam}) the parametrization of ref.~\cite{werner} is to be used; 
strictly speaking, that parameterization holds for the fragmentation into 
$\Lambda + \bar \Lambda$. However, 
in the forward region ($\xi \gsim 0.1$), one expects 
${\cal P}_{\Lambda + \bar\Lambda} \approx \PolL$.  
Furthermore, the parameterization of \cite{werner} assumes $SU(3)$ symmetry:
$D_{\Lambda/u} = D_{\Lambda/d} = D_{\Lambda/s}$. However, the 
polarizing fragmentation functions $\Delta^N D_{\Lambda^\uparrow/q}$
reduce the flavor symmetry to $SU(2)$, 
since $N_{u,d} \neq N_s$ ($N_u=N_d$ was imposed in~\cite{ABDM1} to
reduce the number of fit parameters). But even though 
$\Delta^N D_{\Lambda^\uparrow/s} > |\Delta^N D_{\Lambda^\uparrow/u,d}|$, the 
overall $\Lambda$ polarization in the process under consideration is in 
fact dominated by the valence-like quarks of the proton, not by the strange 
quark. 

The polarizing fragmentation function describes the probability of an
unpolarized quark to fragment into a transversely polarized $\Lambda$. 
Here, no difference is made as to whether the $\Lambda$ is produced directly
or as a secondary particle, for instance as a decay product of heavier 
hyperons
like the $\Sigma^0$ or $\Sigma^{+*}$. This second type is usually expected to
have a depolarizing effect, which means that the degree of polarization is
higher
for the directly produced $\Lambda$'s. The number of directly produced 
$\Lambda$'s is estimated to be roughly 75 \% of the total,
such that the depolarizing effect could be on the order of 30
\%. The polarizing
fragmentation functions of ref.~\cite{ABDM1} thus effectively account for the
depolarizing effect from decays, since they were obtained by a fit to data 
that does not discriminate between direct and decay contributions either. 
  
\begin{figure}[htp]
\centerline{\hbox{\epsfig{figure=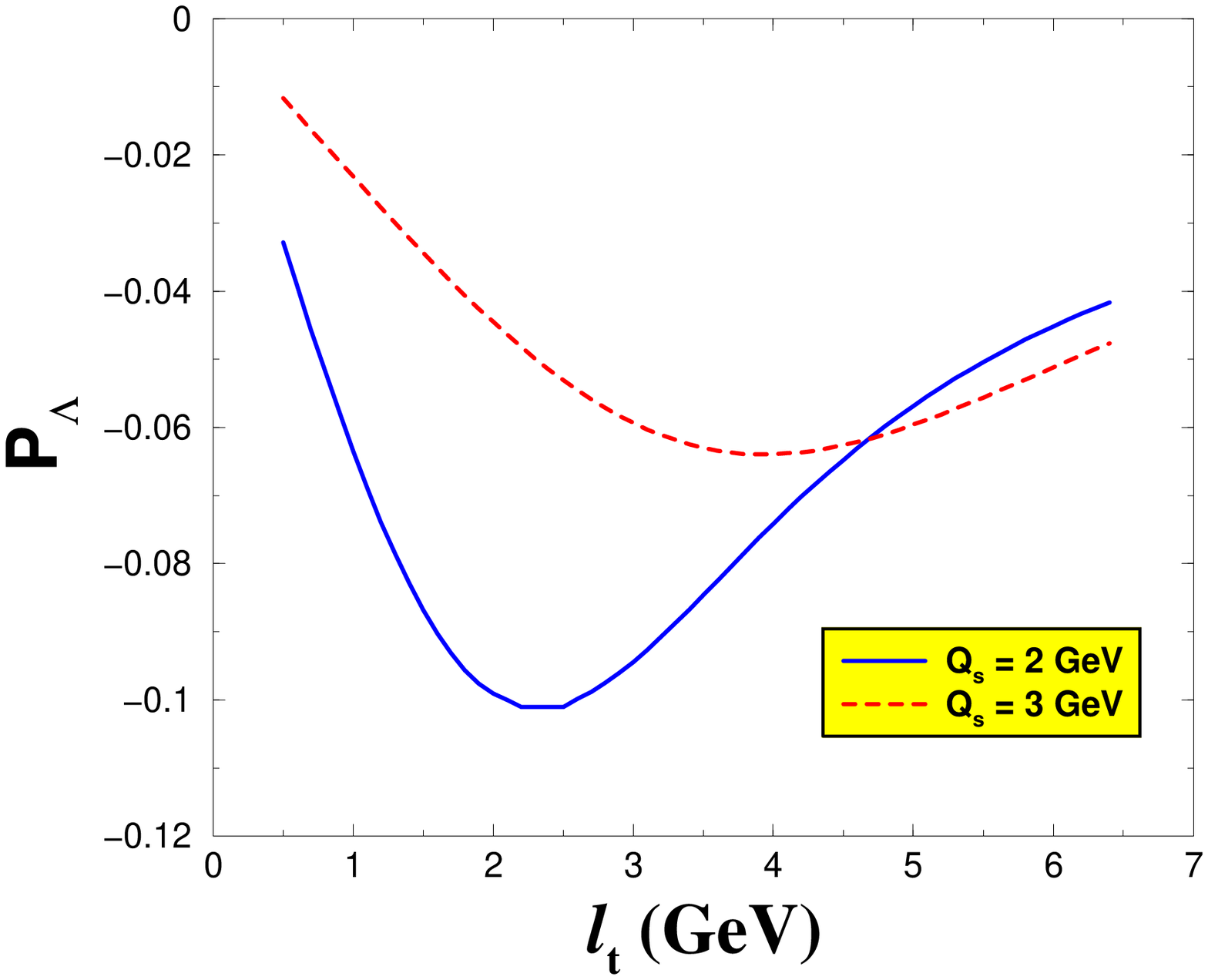,height=7cm}
\epsfig{figure=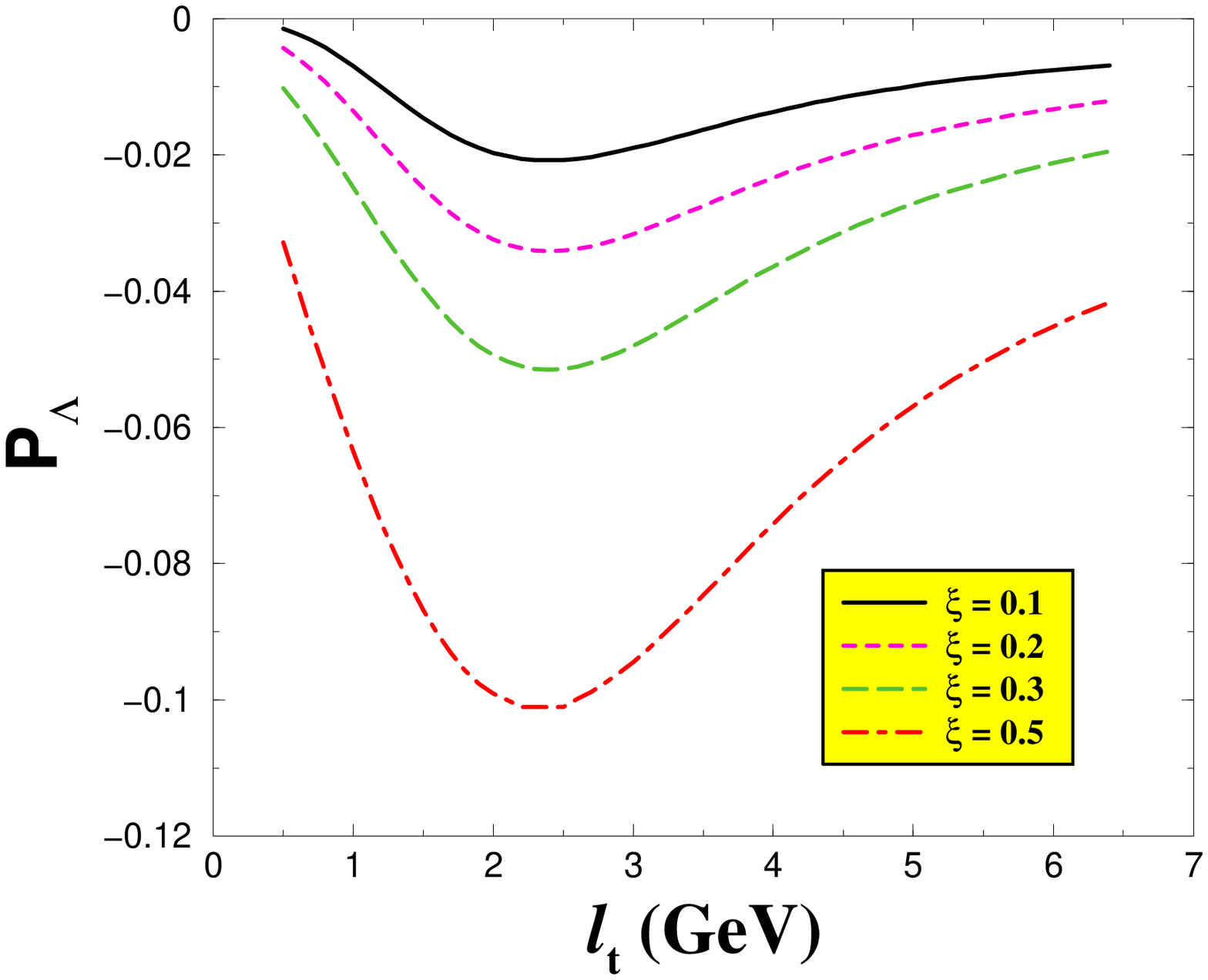,height=7cm}}}
\caption{Transverse momentum distribution of the transverse 
$\Lambda$ polarization. Left: at fixed longitudinal momentum fraction 
$\xi=0.5$ and varying target saturation scale, $Q_s=2,3$~GeV, respectively.
Right: For $Q_s=2$~GeV and various $\xi$.}
  \label{fig1}
\end{figure}
A numerical evaluation of eq.~(\ref{Pol}) 
is shown in Fig.~\ref{fig1}, using the
CTEQ5L LO parton distribution functions for the proton~\cite{cteq}.
Generically, one observes that $\PolL$ is negative (due to the fact that $u$
and $d$ quarks dominate); it first increases with transverse
momentum, then peaks at $l_t\simeq Q_s$, and asymptotically approaches zero
again. 
The fact that $\PolL$ peaks at $l_t\sim Q_s$ has its origin in the
$k_t$-odd nature of the polarizing fragmention function: from eq.~(\ref{Pol}),
$\PolL$ corresponds to the {\em difference} of the cross sections taken
with ``intrinsic'' transverse momentum $k_t^0$ parallel and antiparallel
to the quark transverse momentum $q_t$. Since $k_t^0$ is small, $\PolL$ is
essentially proportional to the {\em derivative} of $d\sigma^{qA}/d^2q_t$,
the differential quark-nucleus cross section, which varies most
rapidly at $q_t\sim Q_s$ (see also eq.~(\ref{Cqt_Qs}) below). Consequently,
$|\PolL|$ exhibits a maximum at such
transverse momentum. This conclusion is independent
of the details of the polarizing fragmentation functions; only the $k_t$-odd
nature and the fact that they are strongly peaked about an average $k_t^0$
matters. In contrast, $k_t$-even distribution and fragmentation
functions only probe the $qA$ cross section itself but not its derivative
with respect to $q_t$.

The behavior of $\PolL(l_t)$ is qualitatively rather different when
the quark cross section is taken at leading twist. In that case, not only is
the magnitude of the polarization larger, but moreover
$\PolL$ in the forward region peaks about small
transverse momentum $\lton1$~GeV. This can be understood by noting
that the derivative of the $q\, A$ cross section at leading twist peaks in
the infrared, contrary to eqs.~(\ref{qAXsec},\ref{Cqt_Qs}).
For a more quantitative evaluation of polarized $\Lambda$ production in the
``dilute regime'' (hadronic collisions far below the unitarity limit,
e.g.\ $pp$ collisions at RHIC) we refer to Ref.~\cite{ABDM1}.

To understand the behavior of eq.~(\ref{Pol}) in more detail, consider
first large transverse momentum, $q_t\gg Q_s$. Here, the last two 
terms of eq.~(\ref{qAXsec}) can be dropped, since they contribute only via a 
$\delta(q_t)$ term. At large transverse momentum the phase 
factor $\exp(i\vec{q}_t \cdot\vec{r}_t)$ in eq.~(\ref{qAXsec}) effectively 
restricts the integral over $d^2r_t$ to the region $r_t\lton1/q_t\ll 1/Q_s$;
the first exponential can then be expanded order 
by order to generate the usual power series in $1/q_t^2$.
The leading and subleading twists are
(see also~\cite{gelis})
\be
C(q_t) =2 \frac{Q_s^2}{q_t^4}\left[1+\frac{4}{\pi}\frac{Q_s^2}{q_t^2}
\log\frac{q_t}{\Lambda} + {\cal O}\left(\frac{Q_s^2}{q_t^2}\right)\right]~.
\label{LT}
\ee
This expression is valid to leading logarithmic accuracy. The first term
corresponds to the perturbative one-gluon $t$-channel exchange contribution 
to $qg \to qg$ scattering \cite{gelis}. 
To leading order in $k^0_t/l_t$, the polarization given in 
eq.~(\ref{Pol}) thus becomes
\be \label{Pol_high_lt}
\PolL(l_t,\xi) = 8
\frac
{\int_\xi^1 dx \, {x} f_{q/p}(x,Q^2) 
 \Delta^N D_{\Lambda^\uparrow/q}\left(\frac{\xi}{x},Q^2\right) 
 x^{-4}
 \left[ 1+\frac{6\xi^2}{\pi x^2} \frac{Q_s^2}{l_t^2}\log\frac{l_t}{\Lambda}
 \right] {k^0_t}/{l_t} 
}
{\int_\xi^1 dx \, {x} f_{q/p}(x,Q^2) 
 D_{\Lambda/q}\left(\frac{\xi}{x},Q^2\right) 
 x^{-4}
 \left[ 1+\frac{4\xi^2}{\pi x^2} \frac{Q_s^2}{l_t^2}\log\frac{l_t}{\Lambda}
 \right]
}~.
\ee
It is known~\cite{ABDM1} that the polarization (for large $l_t$) is a 
higher-twist effect, i.e.\ it is suppressed by powers of the ``intrinsic'' 
transverse momentum at
hadronization, $k^0_t$, over the external momentum scale $l_t$.
Eq.~(\ref{Pol_high_lt}) shows that despite a partial cancellation
the first power-suppressed correction to
the quark-nucleus cross section (the subleading terms in the square brackets)
enhance $\PolL$ at large $l_t$, in agreement with
the behavior at $l_t\gsim5$~GeV in Fig.~\ref{fig1}.

Regarding the scaling of $\PolL$ at the peak,
consider the quark-nucleus cross section
for $q_t\sim Q_s\gg\Lambda$. Again, the last two terms of eq.~(\ref{qAXsec})
can be dropped, while in the leading logarithmic approximation the argument of
the first exponential reads
\be\label{logar}
-\frac{Q_s^2 r_t^2}{4\pi}\log\frac{1}{r_t\Lambda} + {\cal O}(Q_s^2 r_t^2)~.
\ee
The phase factor effectively cuts off the integral at $r_t\sim1/q_t\sim1/Q_s$,
and so $1/r_t\Lambda$ is large. We therefore replace $1/r_t\to Q_s$ in the
argument of the above logarithm, since it is slowly varying and formally
makes the expression well-behaved at large $r_t$. The
remaining integral leads to
\be \label{Cqt_Qs}
C(q_t) \simeq \frac{4\pi^2}{Q_s^2\log\,Q_s/\Lambda}\, \exp\left(-
\frac{\pi q_t^2}{Q_s^2\log\,Q_s/\Lambda}\right)~.
\ee
This approximation reproduces the behavior of the full
expression~(\ref{qAXsec}) about $q_t\sim Q_s$ reasonably well.
Expressions~(\ref{logar},\ref{Cqt_Qs}) are useful only when the cutoff
$\Lambda\ll Q_s$, that is, when color neutrality is enforced on
distance scales of order $1/\Lambda\gg1/Q_s$. If, however, color
neutrality in the target nucleus were to occur on distances of order
$1/Q_s$~\cite{kazu} then $\Lambda\sim Q_s$ and one would have to go beyond
the leading-logarithmic approximation.

From eq.~(\ref{Cqt_Qs}), $\PolL$ is given by
(to leading order in $k_t^0/l_t$)
\be \label{Pol_med_lt}
\PolL(l_t,\xi) = 4 \pi 
\frac{l_t^2}{Q_s^2\log\,Q_s/\Lambda} \frac
{\int_\xi^1 dx \, {x} f_{q/p}(x,Q^2) 
\Delta^N D_{\Lambda^\uparrow/q}\left(\frac{\xi}{x},Q^2\right) \,
\frac{x^2}{\xi^2}\, 
\exp\left(-\frac{\pi l_t^2}{Q_s^2\log\,Q_s/\Lambda}
            \frac{x^2}{\xi^2}\right) {k^0_t}/{l_t}
}
{\int_\xi^1 dx \, {x} f_{q/p}(x,Q^2) 
 D_{\Lambda/q}\left(\frac{\xi}{x},Q^2\right) 
  \exp\left(-\frac{\pi l_t^2}{Q_s^2\log\,Q_s/\Lambda}
            \frac{x^2}{\xi^2}\right)
}~.
\ee
Thus, at the peak
$\PolL$ scales approximately with $1/(Q_s
\surd\log Q_s/\Lambda)$, as indeed seen in Fig.~\ref{fig1}.
The strong dependence on
the target gluon density, as parametrized by $Q_s$, is rather different
from leading-twist perturbation theory.

As mentioned above, there is a related $k_t$-odd effect in processes with one
transversely polarized hadron in the initial state (the Sivers effect), 
which can lead to asymmetries in $p^\uparrow \, p\to \pi \, X$
\cite{Anselmino}, for example. Since at RHIC
polarized proton beams are also available (for recent preliminary 
$p^\uparrow \, p\to \pi \, X$ data from STAR, see Ref.\ \cite{Rakness}), 
one could investigate
the process $p^\uparrow \, A \to \pi \, X$ in the saturation regime.
Similar signatures should arise in that process as for
$p \, A \to \Lambda^\uparrow \, X$ pointed out here. 

In summary, we have studied transverse $\Lambda$ polarization in 
$p \, A$ collisions at high energies and with dense targets. 
The resulting $\Lambda$ polarization is quite different
from that observed in $p \, A$ and $pp$
collisions to date, which presumably
did not probe the saturation regime yet. 
To study the high-density limit, we have performed a
weak coupling analysis of the hard $q\, A$ scattering, determined 
by the saturation momentum $Q_s$, and described the 
unpolarized quark fragmentation into a transversely polarized $\Lambda$
hyperon by the so-called polarizing fragmentation functions.
We observe that the $\Lambda$ polarization peaks
at transverse momentum $\sim Q_s$, where it also
scales approximately as $1/(Q_s \surd\log Q_s/\Lambda)$ and hence is  
collision energy dependent. Moreover, no plateau region for larger transverse
momenta is present. These features are independent of the details of the
polarizing fragmentation functions, but rather occur due to their $k_t$-odd
nature. Similar effects are
expected in the process of $p^\uparrow \, A \to \pi \, X$ in the saturation 
region. Both processes can be studied, in principle, at the BNL-RHIC collider,
and perhaps in the future at the CERN-LHC.

\vspace*{1cm}
{\bf Acknowledgement:}\\
We thank Francois Gelis and Werner Vogelsang for helpful discussions and
for contributing some of their computer codes.
A.D.\ gratefully acknowledges support by the German Minister for
Education and Research (BMBF) and by the U.S.\ Department of Energy
under contract number DE-AC02-98CH10886. The research of D.B.\ has been 
made possible by financial support from the Royal Netherlands Academy of 
Arts and Sciences.

\end{document}